\documentclass{elsart}
\usepackage{graphicx}
\usepackage{amsmath}
\usepackage{amsfonts}
\usepackage{amssymb}
\begin{document}
\begin{frontmatter}
\title
{Does the Holographic Principle\\ determine the Gravitational Interaction?}
\author{F. Canfora and G. Vilasi}
\address
{Istituto Nazionale di Fisica Nucleare, Sezione di Napoli, GC di Salerno\\
Dipartimento di Fisica ''E.R.Caianiello'', Universit\`{a} di
Salerno\\ Via S.Allende, 84081 Baronissi (Salerno), Italy\\
e-mail: canfora@sa.infn.it, vilasi@sa.infn.it}
\begin{abstract}
It is likely that the holographic principle will be a consequence
of the would be theory of quantum gravity. Thus, it is
interesting to try to go in the opposite direction: can the
holographic principle fix the gravitational interaction? It is
shown that the classical gravitational interaction is well inside
the set of potentials allowed by the holographic principle. Computations clarify which role such a principle could have in
lowering the value of the cosmological constant computed in QFT to
the observed one.
\end{abstract}
\end{frontmatter}

\section{Introduction}

\noindent One of the most promising route towards a deeper understanding of
quantum gravity is holography (two detailed reviews are \cite{Bo02}
\cite{AG00}). The pioneering ideas of Bekenstein \cite{Be81}, 't Hooft
\cite{tH93} and Susskind \cite{Su95} shed light on a very peculiar
characteristic of gravitational field which is very likely to survive in the
final theory of quantum gravity. While in Quantum Field Theory (henceforth
QFT), the number of degrees of freedom of a given space-like region is
proportional to the volume of the region itself, if the gravitational effects
are taken into account such a number will appear to be proportional to the
surface of the region. In the elegant framework of \cite{Go83} \cite{Bo99}
\cite{Bo99b} which refined the works \cite{Be81}, \cite{tH93} and \cite{Su95},
the above statement on the degrees of freedom is translated in a covariant
entropy bounds; also a formulation of a causal entropy bound \cite{BV99} is
possible which discloses other important aspects of the above ideas. An
explicit and highly non trivial realization in (super)string theory of the
holographic principle is the AdS/CFT correspondence, first introduced by
Maldacena \cite{Ma97}, in which the role of (super)gravity in decreasing the
number of degrees of freedom, which one would naively expect on QFT grounds,
is manifest. Such a decreasing of the number of degrees of freedom could also
have important consequences, as far as the cosmological constant is concerned,
since the striking disagreement between the cosmological constant computed in
QFT and the observed one of about 120 orders of magnitude (the situation will
be slightly better if supersymmetry is introduced but it remains extremely
unpleasant) is likely to be related to an overcounting of degrees of freedom
in QFT. It is now commonly believed that the property of the number of degrees
of freedom to be proportional to the area of a (suitable defined) surface
enclosing them will be a distinguishing feature of quantum gravity. The
problem is, however, that a theory of quantum gravity is not available yet. A
fruitful point of view which could give even stronger physical basis to the
holographic principle is the following (for a different approach to this
question see, for example, \cite{Pa04}): since the quantum gravity is still
lacking and we cannot deduce from it the holographic principle, can we try to
go in the opposite direction? That means, from the holographic principle is it
possible to deduce (at least) the classical gravitational interaction?

In this paper we move some preliminary steps in this direction. We consider a
closed volume containing a finite but large number of particles interacting
with an \textit{a priori} unknown potential. The question we try to address
is: which form the potential has to have in order for the entropy to be
proportional to the area of the surface which encloses the particles? The
computations are almost entirely classical, nevertheless quantum mechanics has
a very important role in specifying some hypothesis. The results are very
encouraging: the gravitational potential is very likely to be the unique
answer to the above question. This analysis also clarifies the role of the
holographic principle in decreasing the cosmological constant computed in QFT
to the observed value.

\section{The method}

Let us consider a spherical\footnote{The hypothesis of sphericity is not
necessary, it only simplifies computations but the same conclusions can be
reached by dropping it} three dimensional region $S$ with diameter $2\rho$.
Let $N$ denote the number of particles and $u$ be the interaction potential.
Since we will work at a constant density of particles the number of particles
$N$ is proportional to the volume $V(S)$ of $S$:
\begin{equation}
N=c_{1}V(S) \label{densità}%
\end{equation}
$c_{1}$ being a constant with the dimension of an inverse volume proportional
to the density of the gas\footnote{In this section we will not pay much
attention to the physical dimensions of the parameters of the model. A
concrete example of the numerical estimates that one can get will be given in
the last section.}. The classical partition function can be written as
follows:
\begin{align}
Z_{\beta}  &  =\int\overset{i=1,.,N}{\prod}d^{3}p_{(i)}d^{3}q_{(i)}\exp\left[
-\beta\sum H_{(i)}\right]  =\label{partf}\\
&  =\left(  \frac{2\pi}{\beta}\right)  ^{\frac{3N}{2}}\int_{S^{N}}%
\overset{i=1,.,N}{\prod}d^{3}q_{(i)}\exp\left[  -\beta\sum_{\left.
i,j\right\vert i\neq j}u(\overrightarrow{q}_{(i)},\overrightarrow{q}%
_{(j)})\right]  ,\label{part2}\\
H_{(i)}  &  =\frac{1}{2}\left(  \overrightarrow{p}_{(i)}\right)  ^{2}%
+\sum_{\left.  i,j\right\vert i\neq j}u(\overrightarrow{q}_{(i)}%
,\overrightarrow{q}_{(j)}) \label{singham}%
\end{align}
where $\beta$ is a positive real number which can be interpreted as the
inverse temperature, $\overrightarrow{q}_{(i)}$ and $\overrightarrow{p}_{(i)}%
$\ are the position (which is assumed to vary in $S\subset R^{3}$) and momenta
of the $i-$th particle and $u$ is the potential which, as usual, is assumed to
be a binary interaction. The mass of the particles have been set to one. It is
worth to note that, in this \textquotedblright almost
classical\textquotedblright\ model in which quantum mechanics will only enter
through some constraints on the potential and on the parameters of the model,
one can trust computations only above some temperature below which quantum
effects cannot be neglected anymore. We will set the critical value of $\beta$
equal to one (in suitable units): $\beta_{c}=1$ and we will assume that
\[
\beta<\beta_{c}=1.
\]
To be more precise, we could leave $\beta$ unspecified, in this case
computations themselves would tell that $\beta<1$.

The first restriction is to assume $u$ to be a non decreasing function of the
euclidean distance between the $i-$th and the $j-$th particles. The point is
that if $u$ does not increase with the distance then the potential could not
be compatible with bound states and, in such a case, it is very likely that
the holographic bound can not be fulfilled. This fact can be explained as
follows: gravitational interaction is able to lower the effective degrees of
freedom because it is always attractive so that many bound states can be
formed and such bound states behave, in many respects, as single particles. If
the potential decreases with the distance then the particles cannot form bound
states because they prefer to be far apart. The second hypothesis will be to
take the interaction potential depending only on the euclidean distance
between the $i-$th and the $j-$th particles:
\begin{equation}
u=u(\left|  \overrightarrow{q}_{(i)}-\overrightarrow{q}_{(j)}\right|  ).
\label{pot1}%
\end{equation}
This simplifying assumption is not very restrictive, the following
computations can be generalized also to more general ansatz for the
interaction potential. For example, if one assume that
\begin{align}
u  &  =u^{\prime}(\left|  \overrightarrow{q}_{(i)}-\overrightarrow{q}%
_{(j)}\right|  )f(\theta_{(i)},\theta_{(j)},\xi_{(i)},\xi_{(j)}%
)\label{alternpot}\\
0  &  <\alpha_{1}\leq f(\theta_{(i)},\theta_{(j)},\xi_{(i)},\xi_{(j)}%
)\leq\alpha_{2},\nonumber
\end{align}
where $\alpha_{i}$ $i=1,2$ are two positive numerical constants, $\theta
_{(i)}$ are the angular coordinates of $\overrightarrow{q}_{(i)}$ and
$\xi_{(i)}$ are some internal coordinates characterizing the $i-$th particle,
then one will obtain qualitatively the same results provided $u^{\prime}$ is a
non decreasing function of the Euclidean distance between the $i-$th and the
$j-$th particles.

In principle, since we have at our disposal the interaction potential, we can
compute the entropy:
\begin{align}
S_{\beta}  &  =-\beta\partial_{\beta}\ln Z_{\beta}=\frac{c_{1}V(S)I_{\beta
}-\beta\partial_{\beta}I_{\beta}}{I_{\beta}},\label{entr}\\
I_{\beta}  &  =\int_{S^{N}}\overset{i=1,.,N}{\prod}d^{3}q_{(i)}\exp\left[
-\beta\sum_{\left.  i,j\right\vert i\neq j}u(\left\vert \overrightarrow
{q}_{(i)}-\overrightarrow{q}_{(j)}\right\vert )\right]  . \label{integrale}%
\end{align}
where we used Eqs. (\ref{part2}) and (\ref{densità}). Now, the problem is to
find which kind of potentials fulfils the condition
\begin{equation}
S_{\beta}=c_{2}A(\partial S) \label{holo}%
\end{equation}
where $A(\partial S)$ is the area of the boundary $\partial S$ of the region
$S$ and $c_{2}$ is a dimensional constant to be specified later on. The above
holographic requirement can be thought as a differential equation for the
integral $I_{\beta}$ defined in Eq. (\ref{integrale}):
\begin{equation}
\frac{c_{1}V(S)I_{\beta}-\beta\partial_{\beta}I_{\beta}}{I_{\beta}}%
=c_{2}A(\partial S). \label{inte2}%
\end{equation}
The above equation can be easily solved as follows
\begin{equation}
I_{\beta}=k\beta^{\left[  c_{1}V(S)-c_{2}A(\partial S)\right]  }
\label{tempdip}%
\end{equation}
where $k$ is an integration constant. Thus, we have to deduce which conditions
on the potential (remember that $I_{\beta}$ depends on the potential through
Eq. (\ref{integrale})) the above form of $I_{\beta}$ implies. At a first
glance, it seems unlikely for Eq. (\ref{tempdip}) to be able to severely
constraint the interaction potential since, roughly speaking, many functions
can have the same integral. In fact, by taking into account that Eq.
(\ref{integrale}) should hold for any volume not too small (in a sense which
will be clarified later on) contained in $S$ and containing all the particles
as well as the line segment joining them, it will be possible to find powerful
restrictions on the form of $u$.

\subsection{Constraints on the parameters}

The role of quantum mechanics will be to translate the fact that this model
can be trusted only in a classical regime into inequalities between the
parameters. The first constraint is related to the range of the argument
$\left\vert \overrightarrow{q}_{(i)}-\overrightarrow{q}_{(j)}\right\vert $ of
the potential which cannot be to small
\begin{equation}
\frac{\rho}{l_{0}}\leq\left\vert \overrightarrow{q}_{(i)}-\overrightarrow
{q}_{(j)}\right\vert \leq2\rho,\quad l_{0}\gg1 \label{co1}%
\end{equation}
$l_{0}$ being a large positive number to be fixed. The above inequality is
related to the fact that the classical form of an interaction potential is
valuable only above a certain length scale below which some unknown quantum
effects set in. Thus, the number $l_{0}$ is related not only to the arising of
quantum effects but also to the potential.

The second constraint is purely quantum mechanical in nature and is only
related to holography: it deals with the minimum length below which we cannot
resolve anymore distinct particles inside $S$ due to quantum effects. A
reasonable order of magnitude for such a scale, which is suggested by the
holographic principle is
\begin{equation}
\rho_{\min}=\frac{a}{\rho^{2}}. \label{minradius}%
\end{equation}
where $a$ is a constant with the dimension of a volume. The point is that,
according to the holographic bounds, $\rho^{2}$ should be an upper bound for
the total entropy of $S$. The more the entropy, the more different quantum
states are available inside $S$; the more different quantum states, the easier
will be to distinguish distinct particles. In other words, if inside $S$ only
few quantum states are available, for example only one, then it will be
impossible to distinguish distinct particles with physical measurements since
they will be in a single quantum entangled states which would be destroyed by
an external action\footnote{Another way to argue that the above order of
magnitude is reasonable is to consider a lenght scale of order $\rho/N$ where
$N$ is the total number of particles. Because of Eq. (\ref{densità}), one can
assume $N\sim\rho^{3}$ so that the minimum radius below which we cannot
resolve two particles anymore will be $\varpropto\rho^{-2}$. However, the
first argument since it is ''purely holographic''\ and so does not refer to
Eq. (\ref{densità}).}.

It is worth to note that this hypothesis is rather natural since, as it is
well known, in the limit of large quantum numbers the quantum effects can be
neglected (see, for example, \cite{Ya82}, \cite{BB03}).

\section{The model}

The first problem, in trying to understand which are the possible forms of the
interaction potential compatible with Eq. (\ref{tempdip}), is that it is a
rather unusual constraint. In order to analyze the Eq. (\ref{tempdip}), it is
possible to develop an approximate method which gives very accurate results
when the number of particles is large enough.

The geometrical setting is the following. Let us denote with $T_{ij}$ a
cylinder connecting the particle $i$ and the particle $j$. Let the diameter of
the section of $T_{ij}$ be equal to $2\rho_{\min}$ (which is defined in Eq.
(\ref{minradius})) in such a way that inside the cylinder $T_{ij}$ there are
only the particle $i$ and the particle $j$. Remember that, in any case, the
distance between every pair of particles has to be greater than $\rho/l_{0}$
and smaller than $2\rho$ so that the minimum height for the cylinder will be
$\rho/l_{0}$ and the maximum height will be $2\rho$. Thus, the diameter of the
section of the cylinders is related to the smallest geometric scale which is
resolvable with physical measurements, while the minimum height is related to
the minimum scale below which the potential cannot work well anymore.

Hence, we have the following inequalities for the volumes ($V$) and the areas
($A$) of the cylinders $T_{ij}$:%
\begin{equation}
\pi a^{2}\frac{1}{l_{0}\rho^{3}}\leq V(T_{ij})\leq2\pi a^{2}\frac{1}{\rho^{3}%
},\quad\pi a\frac{1}{l_{0}\rho}\leq A(\partial T_{ij})\leq\pi a\frac{1}{\rho}.
\label{vol1}%
\end{equation}

Extending the above inequalities to the union
\begin{equation}
\Gamma=\bigcup_{i\neq j}^{1\leq i,j\leq N}T_{ij} \label{vol3}%
\end{equation}
of all $\left(  N-1\right)  N/2$ cylinders, we obtain
\begin{align}
V_{\min}\left(  \Gamma\right)   &  =\frac{\left(  N-1\right)  N}{2}\pi
a^{2}\frac{1}{l_{0}\rho^{3}}\leq V(\Gamma)\leq\frac{\left(  N-1\right)  N}%
{2}2\pi a^{2}\frac{1}{\rho^{3}}=V_{\max}\left(  \Gamma\right) \label{vol2}\\
A_{\min}\left(  \partial\Gamma\right)   &  =\left(  N-1\right)  N\pi a\frac
{1}{l_{0}\rho}\leq A(\partial\Gamma)\leq2\left(  N-1\right)  N\pi a\frac
{1}{\rho}=A_{\max}\left(  \partial\Gamma\right)  . \label{area1}%
\end{align}

$\Gamma$\ is a sort of fat graph whose vertices are the particles and whose
bold links are the cylinders. The holographic principle (\ref{holo}) has to
hold for any connected subset of $S$ containing all the particles and, in
particular, it has to hold for $\Gamma$ defined in Eq. (\ref{vol3}). Since $u$
is supposed to be a non decreasing function of the distance between $i$ and
$j$ one gets:
\begin{equation}
-\beta u(\frac{\rho}{l_{0}})\geq-\beta u(\left\vert \overrightarrow{q}%
_{(i)}-\overrightarrow{q}_{(j)}\right\vert )\geq-\beta u(2\rho). \label{vol7}%
\end{equation}
On the other hand, Eq. (\ref{tempdip}) must hold, so that from Eqs.
(\ref{vol2}), (\ref{vol7}) and (\ref{integrale}) it follows (remember that
$0<\beta<1$)
\begin{align}
k\beta^{\left[  c_{1}V_{\max}\left(  \Gamma\right)  -c_{2}A_{\min}\left(
\partial\Gamma\right)  \right]  }  &  \leq k\beta^{\left[  c_{1}%
V(\Gamma)-c_{2}A(\partial\Gamma)\right]  }=I_{\beta}\leq\exp\left[
-\frac{\left(  N-1\right)  N}{2}\beta u(\frac{\rho}{l_{0}})\right]
\times\nonumber\\
V(\Gamma)^{N}  &  \leq\frac{\left[  \left(  N-1\right)  N\pi a^{2}\right]
^{N}}{\rho^{3N}}\exp\left[  -\frac{\left(  N-1\right)  N}{2}\beta u(\frac
{\rho}{l_{0}})\right]  ,\label{vol11}\\
k\beta^{\left[  cV_{\min}\left(  \Gamma\right)  -c_{2}A_{\max}\left(
\partial\Gamma\right)  \right]  }  &  \geq k\beta^{\left[  c_{1}%
Vol(\Gamma)-c_{2}Area(\partial\Gamma)\right]  }=I_{\beta}\geq\exp\left[
-\frac{\left(  N-1\right)  N}{2}\beta u(2\rho)\right]  \times\nonumber\\
V(\Gamma)^{N}  &  \geq\frac{\left[  \frac{\left(  N-1\right)  N}{2}\frac{\pi
a^{2}}{l_{0}}\right]  ^{N}}{\rho^{3N}}\exp\left[  -\frac{\left(  N-1\right)
N}{2}\beta u(2\rho)\right]  , \label{vol12}%
\end{align}
where $V_{\max}\left(  \Gamma\right)  $, $V_{\min}\left(  \Gamma\right)  $,
$A_{\min}\left(  \partial\Gamma\right)  $ and $A_{\max}\left(  \partial
\Gamma\right)  $\ have been defined in Eqs. (\ref{vol2}) and (\ref{area1}).
From Eqs. (\ref{vol11}) and (\ref{vol12}) it follows
\begin{align}
c_{1}\left(  \ln\beta\right)  \left(  N-1\right)  N\frac{\pi a^{2}}{\rho^{3}%
}+\frac{\left(  N-1\right)  N}{2}\beta u(\frac{\rho}{l_{0}})  &
\leq\nonumber\\
c_{2}\left(  \ln\beta\right)  \left(  N-1\right)  N\frac{\pi a}{l_{0}\rho
}+L_{\max}-3N\ln\rho & \label{vol14}\\
c_{1}\left(  \ln\beta\right)  \left(  N-1\right)  N\frac{\pi a^{2}}{2l_{0}%
\rho^{3}}+\frac{\left(  N-1\right)  N}{2}\beta u(2\rho)  &  \geq\nonumber\\
2c_{2}\left(  \ln\beta\right)  \left(  N-1\right)  N\frac{\pi a}{\rho}%
+L_{\min}-3N\ln\rho &  , \label{vol15}%
\end{align}%
\begin{equation}
L_{\max}=\ln\left\{  \frac{\left[  \left(  N-1\right)  N\pi a^{2}\right]
^{N}}{k}\right\}  ,\quad L_{\min}=\ln\left\{  \frac{\left[  \left(
N-1\right)  N\frac{\pi a^{2}}{2l_{0}}\right]  ^{N}}{k}\right\}  .
\label{vol16b}%
\end{equation}
From Eqs. (\ref{vol14}) and (\ref{vol15}), upon a trivial rescaling of the
argument of $u$, one gets the following bounds:
\begin{align}
u(r)  &  \leq\left(  \frac{c_{2}\left(  \ln\beta\right)  2\pi a}{\beta
l_{0}^{2}}\right)  \frac{1}{r}-\left(  \frac{c_{1}\left(  \ln\beta\right)
2\pi a^{2}}{\beta l_{0}^{3}}\right)  \frac{1}{r^{3}}+\nonumber\\
&  -\frac{6N}{\beta N(N-1)}\ln(l_{0}x)+\Lambda_{\max}\label{vol21}\\
u(r)  &  \geq\left(  \frac{c_{2}\left(  \ln\beta\right)  8\pi a}{\beta
}\right)  \frac{1}{r}-\left(  \frac{c_{1}\left(  \ln\beta\right)  8\pi a^{2}%
}{\beta l_{0}}\right)  \frac{1}{r^{3}}+\nonumber\\
&  -\frac{6N}{\beta N(N-1)}\ln(\frac{r}{2})+\Lambda_{\min}\label{vol22}\\
\Lambda_{\max}  &  =\frac{2}{\beta N(N-1)}L_{\max},\ \Lambda_{\min}=\frac
{2}{\beta N(N-1)}L_{\min} \label{vol23}%
\end{align}
where $L_{\max}$ and $L_{\min}$\ have been defined in Eq. (\ref{vol16b}).
Eventually, by taking the limit for $N\rightarrow\infty$, the result is
\begin{align}
-\frac{G_{\max}}{r}+\frac{C_{\max}}{r^{3}}  &  \leq u(r)\leq-\frac{G_{\min}%
}{r}+\frac{C_{\min}}{r^{3}}\label{vol21bis}\\
G_{\max}  &  =\left\vert \frac{c_{2}\left(  \ln\beta\right)  8\pi a}{\beta
}\right\vert ,\quad G_{\min}=\left\vert \frac{c_{2}\left(  \ln\beta\right)
2\pi a}{\beta l_{0}^{2}}\right\vert ,\label{vol24}\\
C_{\max}  &  =\left\vert \frac{c_{1}\left(  \ln\beta\right)  8\pi a^{2}}{\beta
l_{0}}\right\vert ,\quad C_{\min}=\left\vert \frac{c_{1}\left(  \ln
\beta\right)  2\pi a^{2}}{\beta l_{0}^{3}}\right\vert , \label{vol25}%
\end{align}
where
\[
r>3\frac{ac_{1}}{l_{0}c_{2}},
\]
otherwise the hypothesis that $u$ is a non decreasing function would be violated.

\section{Physical implications}

First of all, the inequalities (\ref{vol21}) and (\ref{vol22}) will be
consistent only when $\beta<1$ in such a way that the terms in round brackets
multiplying $1/r$ are negative. We observe from Eq. (\ref{vol21bis}) that, for
$r$ big enough, the terms of order $1/r^{3}$ are negligible and $u$ is
constrained to lie between two Newtonian-like potentials. It is remarkable
that the Newtonian terms only depends on purely \textquotedblright
holographic\textquotedblright\ constants. That is, the constants in the round
brackets multiplying $1/r$ only depend on $\beta$, $a$ and $c_{2}$.

This is a highly non trivial self-consistency check of the fact that the
holographic principle can really fix the form of the gravitational
interaction: the constant $c_{2}$ is the constant multiplying the area of $S$
in the holographic constraint (\ref{holo}) and also $a$ enters in the
definition of the holographic length scale (\ref{minradius}). Thus, the
Newtonian part is constrained only by the holographic principle (as it should
be), while the $1/r^{3}$ terms, which do not vanish for $N\rightarrow\infty$,
contain the proportionality constant between the volume and the entropy in the
limit of vanishing potential.

Another important aspect of the above model is the following. In our universe,
the number of particles is very large but, actually, it is not infinite. For
this reason, there is the possibility of a very small constant to be added to
the potential. The inequalities (\ref{vol21}) and (\ref{vol22}) tell us that
the magnitude of such a constant should lie between $\Lambda_{\min}$\ and
$\Lambda_{\max}$\ defined in Eq. (\ref{vol23}). It is very interesting to note
that in $\Lambda_{\min}$\ and $\Lambda_{\max}$, besides a geometrical factor,
it appears a factor $\left(  \log N\right)  /N$ which (being $N\sim10^{120}$ a
reasonable estimate of the total number of particles in the universe) is of
order $10^{-118}$. Such a small factor could resolve the problem of the too
large value QFT-computed cosmological constant: if the coupling of gravity
with quantum fields is such that holography is preserved, then the quantized
version of the holographic constraint (\ref{holo}) could provide the right
factor to suppress the ''bare''\ QFT cosmological constant. The holographic
principle, by properly taking into account the effects of gravity on the
degrees of freedom, could renormalize the ''bare''\ QFT cosmological constant
with a very small factor. An intuitive explanation of this fact is that QFT
counts as distinct pair of degrees of freedom which, in fact, coupled by
gravity behave as single degrees of freedom and should, therefore, not be overcounted.

A more formal way to understand this fact, which also clarify that the above
argument could also hold at a quantum level, is the following. One of the main
reason behind the fact that in QFT the computed cosmological constant is too
large is that in the path integral approach one performs the diagrammatic
expansion starting from classical vacua. On the other hand, classical vacua
(that is, stable solutions of the classical equations of motions) are
invariant if we add to the Hamiltonian a constant term so that they cannot
provide us with an energy scale. The energy scale (''zero point energy'') of
QFT is a purely quantum effects which, when renormalized, is of the order of
the $UV$ cutoff and, therefore, too large. If the coupling with gravity is
standard (via Lagrangian or Hamiltonian), then the above problem, at a first
glance, will be only slightly softened for the same reasons as above. Instead,
let us consider an equation like
\begin{align}
-\beta\partial_{\beta}\ln Z_{\beta}  &  =\frac{1}{4L_{P}^{2}}A(\partial
S)\label{quholo}\\
Z_{\beta}  &  =\int\left.  \left[  D\Phi^{a}Dg_{\mu\nu}D\pi_{\mu\nu}\right]
\right|  _{B.C.\rightarrow\partial S}\exp\left[  -\beta\int_{S}H(\Phi
^{a},g_{\mu\nu},\pi_{\mu\nu})dS\right]  \label{qupart}%
\end{align}
where $L_{P}$ is the Planck length, $D$ is the standard notation for the
''path''\ integration, $\Phi^{a}$\ is a collective symbol to denote the set of
quantum fields and their conjugated momenta besides gravity (whose phase space
variables are denoted with $g_{\mu\nu}$ and $\pi_{\mu\nu}$) and the symbol
$B.C.\rightarrow\partial S$ means that we must impose suitable boundary
condition on the fields as they approach the boundary of $S$. Gravity should
be introduced by requiring that the metric needed to compute the curved
partition function (\ref{qupart}) is such that Eq. (\ref{quholo}) is
fulfilled. Although to solve explicitly this problem seems to be a rather
hopeless task, a very interesting result is now apparent. We are not free
anymore to add an arbitrary constant to the density of Hamiltonian, since Eq.
(\ref{quholo}) is not invariant under the transformation
\begin{align}
H  &  \rightarrow H+\Lambda_{0}\label{bresca}\\
\Lambda_{0}  &  =const,\nonumber
\end{align}
in other words, an holographic equation like Eq. (\ref{quholo}) set a scale
for the zero point energy. Even if a rigorous quantum computation is required
to obtain the exact value of $\Lambda_{0}$, it is nevertheless interesting to
note that Eq. (\ref{quholo}) tells that $\Lambda_{0}$ should be positive and
of the order (such an estimate could be improved by a careful quantum
computation)
\begin{align}
\Lambda_{0}  &  \sim H(0,g_{\mu\nu}=\eta_{\mu\nu},0)\sim\frac{1}{4\beta
L_{P}^{2}}\frac{A(\partial S)}{V(S)}\sim M_{P}^{4}\left(  \frac{k_{b}T}{M_{P}%
}\right)  \left(  \frac{L_{P}}{L_{U}}\right) \label{coco}\\
L_{U}  &  =\frac{V(S)}{A(\partial S)},\nonumber
\end{align}
where $M_{P}$ is the Planck mass and $L_{U}$ should be a length measuring the
size of the space-time region past causally connected today. If we take as
$L_{U}$ the Hubble radius today and as $T$ the cosmic temperature today we get%
\[
\Lambda_{0}\sim M_{P}^{4}\times10^{-95}.
\]
This result is very promising if we consider that, in this computation (even
if QFT has been taken into account in a rather rough way) it has been achieved
a striking reduction (of, at least, 30 orders of magnitude) of the values of
the cosmological constant which are usually obtained in different context
(QFT, SUSY QFT, SUGRA and so on) without introducing SUSY, extra dimensions or
fine-tuning of any kind. Eq. (\ref{coco}) also suggests that, in the past, the
cosmological ''constant''\ should had been (much) greater. This implies a
modification of Einstein equations. The necessity of some kind of modification
of the Einstein equations is, today, widely recognized due, for example, to
the experimental data on the accelerated expansion. The best explanation of
the actual experimental data would be a ''varying''\ cosmological constant
much higher in the past than now (see, for example, \cite{Sa02}). On the other
hand, there is not in the standard approaches a natural way to achieve this
goal. Usually, non minimally coupled scalar fields (whose physical origin is,
however, unknown) are introduced in order to imitate the behavior of a
''varying''\ $\Lambda$\ term (see, for example, \cite{Sa02}). In fact, the
holographic principle can give rise to results in a much better agreement with
experimental data providing, at the same time, with a more natural explanation
of a $\Lambda$ decreasing with time.

\section{Conclusion}

In this paper an alternative point of view to analyze the relations between
gravity and holography has been proposed. It is commonly believed that the
holographic principle will be a corollary of the final theory of quantum
gravity which, on the other hand, is not available yet. For this reason, it is
interesting to try to go in the opposite direction: is it possible to deduce
the gravitational interaction from the holographic principle? Here, the first
preliminary steps in this direction have been performed. It has been shown
that the classical gravitational potential belongs to the rather narrow region
allowed by the holographic principle. It has been clarified the physical
mechanism responsible for the smallness of the observed cosmological constant
with respect the one computed in QFT: QFT overcounts pairs of degrees of
freedom which, in fact, coupled by gravity, behave as single degrees of
freedom and should not be overcounted. The holographic principle could face
with this overcounting thanks to a very small multiplying factor. Even if the
computation are entirely classical, it has been argued that the same could
hold at a quantum level; a promising estimate of the order of magnitude of the
cosmological constant has also been obtained. The reason for this is that Eq.
(\ref{quholo}) is not invariant when we add to the density of Hamiltonian a
constant. In other words, the holographic principle encodes a natural scale
for the cosmological constant which is likely to survive also at a quantum level.

\end{document}